\documentclass[a4paper,twocolumn,11pt]{quantumarticle}
\pdfoutput=1
\usepackage[utf8]{inputenc}
\usepackage[english]{babel}
\usepackage[T1]{fontenc}
\usepackage{amsmath}
\usepackage{hyperref}

\usepackage{lipsum}
\usepackage{tikz}
\usepackage{amsmath}
\usetikzlibrary{shapes.geometric, arrows}
\usepackage{physics}
\usepackage{braket}
\usepackage{bm}
\usepackage{appendix}
\usepackage{float}
\usepackage{amssymb}
\usepackage{amsthm}
\newtheorem{theorem}{Theorem} 
 
\usepackage[normalem]{ulem}

\usepackage{xcolor}

\newcommand{\cN}{\mathcal{N}}
\newcommand{\cR}{\mathcal{R}}

\newcommand{\cI}{\mathcal{I}}

\newcommand{\T}{{\rm{T}}}

\newtheorem{proposition}[theorem]{Proposition}
\newtheorem{corollary}[theorem]{Corollary}

\begin{document}

\title{Universal Optimization and Tighter Fidelity Bounds for Approximate Quantum Error Correction}

\author{Jing Wu}
\affiliation{Fermi National Accelerator Laboratory, Batavia, IL 60510, USA}
\orcid{0000-0002-4946-0732}

\author{Michele Grossi}
\affiliation{European Organization for Nuclear Research (CERN), CH-1211 Geneva 23, Switzerland}
\author{Do\~{g}a Murat K\"{u}rk\c{c}\"{u}o\~{g}lu}
\affiliation{Fermi National Accelerator Laboratory, Batavia, IL 60510, USA}
\author{Silvia Zorzetti}
\affiliation{Fermi National Accelerator Laboratory, Batavia, IL 60510, USA}
\orcid{0000-0002-3208-3387}
\email{zorzetti@fnal.gov}
\maketitle

\begin{abstract}
Approximate quantum error correction (AQEC) extends the framework of discrete- and continuous-variable quantum error correction beyond the Knill-Laflamme (KL) conditions, where the recovery performance is quantified by entanglement fidelity. Recent studies have enabled efficient evaluation of near-optimal entanglement fidelity using transpose-channel recovery. Yet, determining the global optimal recovery map and its entanglement fidelity for general codes beyond the KL conditions remains a major computational challenge. Direct optimization becomes prohibitive as the number of noise Kraus operators grows rapidly with system size, and existing approaches lack rigorous guarantees for reducing this optimization to a tractable dimension.
Here, we derive an explicit characterization of the optimal environmental state of complement channel, which transforms the optimization over recovery channels into an equivalent optimization over quotient unitaries. For a broader class of codes that satisfy only the orthogonality part of the KL conditions, we show that the optimal recovery map admits an explicit analytical form. Building on this form, we derive novel rigorous lower bounds of entanglement fidelity that strictly improve upon the transpose-recovery bound.
We further develop a novel recovery strategy based on principle components, and derive a rigorous bound on the error introduced by noise truncation.
Our approach enables efficient searches for approximate recovery maps for AQEC codes, avoiding the need to optimize over the full Kraus-operator space.

\end{abstract}
\section{Introduction}
Approximate quantum error correction (AQEC) has emerged as a unifying theoretical framework across diverse frontiers of modern physics \cite{Schumacher1997AQEC, beny2010general}. In the context of quantum information processing, AQEC extends exact error correction~\cite{knill1997theory}, providing realistic optimization strategies to safeguard discrete-variable (DV)~\cite{shor1995scheme} and continuous-variable (CV)~\cite{gottesman2001encoding, albert2018performance} quantum processors against hardware decoherence. Beyond quantum computing, AQEC provides a foundational language to describe diverse physical phenomena. Under the anti-de Sitter/conformal field theory (AdS/CFT) correspondence, it frames bulk spacetime reconstruction from boundary degrees of freedom as an isometric code embedding~\cite{almheiri2015bulk, pastawski2015holographic, cotler2019approximate}. In condensed matter physics, AQEC models the resilience and phase transitions of topological phases against thermal anyonic excitations at finite temperatures~\cite{dennis2002topological, alicki2009thermal}. Furthermore, it quantifies the ultimate limits of work extraction and entropy production under strong-coupling quantum thermodynamics~\cite{berta2017thermal}.
Consequently, identifying the optimal recovery map for an AQEC code across these diverse physical scenarios has garnered broad interest, yet it remains a fundamental challenge, particularly for multi-qubit DV architectures and CV systems requiring large Fock-space truncation.

Traditional quantum error correction (QEC) codes rely on exact conditions, i.e. the Knill-Laflamme (KL) conditions~\cite{Knill1998Threshold, Knill1997KLCondition}, under which the Petz recovery map (transpose channel)~\cite{Petz1986Recovery, Petz1988Sufficiency} ensures perfect recovery. 
However, for realistic noise processes—such as thermal loss and amplitude damping—codes that exactly satisfy the KL conditions are generally unattainable within experimentally feasible encoding spaces. These physical realities forces a paradigm shift to AQEC, where the Petz recovery map is no longer optimal~\cite{Junge2018Universal}, and finding the recovery map that maximizes the entanglement fidelity is conventionally formulated as a semidefinite program (SDP)~\cite{fletcher2007optimum}. While SDP guarantees global optimality for small systems, its computational complexity scales prohibitively with the number of Kraus operators. For advanced CV and DV codes, as well as the AQEC frameworks underlying these diverse physical phenomena, the number of Kraus operators routinely grows into the thousands as system size scales up. In this high-dimensional regime, standard SDP encounters a severe numerical bottleneck, rendering high-precision optimization an intractable task.

In this work, we overcome this bottleneck by introducing a dual representation for the maximization of entanglement fidelity. This dual perspective enables us to establish analytical lower bounds on entanglement fidelity that are significantly sharper than the near-optimal fidelity bounds derived from the transpose channel \cite{Near-optimal}. Within this representation, we reformulate the optimization problem for the recovery map as a constrained search over a quotient manifold and integrate a principal component analysis (PCA)-based framework. We demonstrate that the PCA-based framework is sufficiently precise to guarantee tight accuracy bounds while drastically reducing the computational overhead in high-dimensional systems. We validate the versatility of our framework by applying it to both CV and DV systems. Specifically, we evaluate the entanglement fidelity of the Gottesman-Kitaev-Preskill (GKP) code~\cite{Gottesman2001GKP} subject to thermal loss, as well as the Shor nine-qubit code under amplitude damping. Our framework offers a robust and scalable toolset for AQEC development, facilitating rapid yet highly accurate evaluation of error-correcting codes embedded in high-dimensional error spaces.

\section{Entanglement fidelity in dual representation}
Given a noisy channel $\mathcal{N}$, QEC aims to find a recovery map $\mathcal{R}$ such that the composite channel $\mathcal{R} \circ \mathcal{N}$ approximates the identity $\mathcal{I}$ on the encoded subspace $\mathcal{H}_{\text{enc}}$. While code performance can be evaluated by the average state fidelity $F_{\text{avg}} = \int_{\mathcal{H}_{\text{enc}}} d\psi \bra{\psi} (\mathcal{R} \circ \mathcal{N})(\ketbra{\psi}) \ket{\psi}$, it is more advantageous to employ the entanglement fidelity defined by quantum state fidelity
\begin{align}
F_e(\mathcal{R}\circ\mathcal{N},\cI) &= F\left[ (\mathcal{R}\circ \mathcal{N}\otimes \mathcal{I}_{S'}) (\ketbra{\Phi}), \right. \notag\\
&\quad \left. (\cI \otimes \cI_{S'})(\ketbra{\Phi}) \right],
\end{align}
where $\ket{\Phi} = \frac{1}{\sqrt{d_L}} \sum_{i=1}^{d_L} \ket{i,i}_{SS'}$ is the maximally entangled state between the system $S$ (with orthonormal basis $\{\ket{i}_S\}$ spanning $\mathcal{H}_{\text{enc}}$) and a reference $S'$. These two metrics are connected via the Horodecki-Nielsen identity~\cite{Horodecki1999Link, Nielsen1999SimpleFormula}
\begin{equation}
F_{\text{avg}} = \frac{d_L F_e + 1}{d_L + 1}.
\end{equation}
Consequently, maximizing $F_e$ guarantees the optimal preservation of the entire code subspace. Let define the environment-reference state $\rho \equiv (\mathcal{N}^C \otimes \mathcal{I}_{S'})(\ketbra{\Phi})$, where $\mathcal{N}^C$ is the complementary channel. For a quantum channel $\mathcal{N}$ characterized by a set of Kraus operators $\{E_k\}_{k=1}^{d_E}$, the state $\rho$ is explicitly given by
\begin{equation}
\rho = \frac{1}{d_L} \sum_{i,j=1}^{d_L} \; \sum_{k,l=1}^{d_E} \bra{j} E_l^\dagger E_k \ket{i} \ketbra{k}{l}_E \otimes \ketbra{i}{j}_{S'}.
\end{equation}
By virtue of the AQEC duality theorem~\cite{Beny2010AQEC}, finding the optimal recovery map $\mathcal{R}$ is dual to finding a state-preparation channel $\mathcal{R}'$ that maximizes the entanglement fidelity of the complementary channel $\mathcal{N}^C$, satisfying $\max_{\mathcal{R}} F_e(\mathcal{R}\circ\mathcal{N}, \mathcal{I}) = \max_{\mathcal{R}'} F_e(\mathcal{N}^C, \mathcal{R}')$. 
Since 
\begin{align}
    \mathcal{R}'(\ketbra{\Phi}) = \sigma_E \otimes I_{S'}/d_L,
\end{align}
where $\sigma_E$ is a normalized environment state, the optimization reduces to
\begin{equation}
\max_{\mathcal{R}} F_e(\mathcal{R}\circ\mathcal{N}, \mathcal{I}) = \max_{\sigma_E} F\left( \rho, \sigma_E \otimes \frac{I_{S'}}{d_L} \right).
\label{eq:dual_F}
\end{equation}
Utilizing Uhlmann's theorem~\cite{Uhlmann1976}, the dual maximization in Eq.~\eqref{eq:dual_F} admits an analytical solution for $\sigma_E$ in Uhlmann's theorem (proved in Appendix~\ref{app1}). The recovery map $\mathcal{R}$ is thus obtained in closed form via $U$, yielding our central result:
\begin{theorem}[Entanglement fidelity in dual representation]
The maximum entanglement fidelity is given by
\begin{equation}
\max_{\mathcal{R}} F_e(\mathcal{R} \circ \mathcal{N}, \mathcal{I}) = \max_{U \in \mathcal{U}(d_Ed_L)} \frac{1}{d_L} \left \lVert \Tr_{S'} (U\sqrt{\rho}) \right \rVert_F^2,
\label{eq:main}
\end{equation}
where $U$ is a unitary on $ES'$ and $\lVert\cdot\rVert_F$ is the Frobenius norm. The recovery map is given by
\begin{equation}
\mathcal{R}(\rho_S) = \Tr_A \left( U_{\mathcal{R}} (\rho_S \otimes \ketbra{0}_A) U_{\mathcal{R}}^\dagger \right),
\label{eq:recovery_map}
\end{equation}
where the recovery unitary $U_{\mathcal{R}}^\T = \left( |X|X^+ \otimes I \right) U$, with the reduced operator denoting $X \equiv \Tr_{S'}(U\sqrt{\rho})$, and $(\cdot)^+$ is the Moore-Penrose pseudoinverse.
\end{theorem}
Similar to the conventional SDP optimization that operates on $d_E d_L \times d_E d_L$ complex matrices, the formulation here is defined over the unitary manifold $\mathcal{U}(d_E d_L)$. When we relax the constraint to $\|U\| \le 1$, the problem becomes a convex optimization problem that yields a unique maximal value on the boundary $\|U\| = 1$.
For exact QEC, the KL conditions are formulated as $\rho = \sigma_E \otimes I_{S'}/d_L$. In this idealized scenario, it is straightforward to verify that the identity matrix $U = I$ serves as the optimal unitary, yielding unit entanglement fidelity. 
In contrast, for AQEC, the KL conditions are no longer strictly satisfied. The objective shifts to identifying an optimal disentangling unitary $U$ in Eq.~\eqref{eq:main} that transforms the state $\rho$ into a form as close as possible to the ideal product state $\sigma_E \otimes I_{S'}/d_L$ in the dual representation. 

In general, a universal closed-form solution for the optimal unitary $U$ that maximizes the entanglement fidelity in Eq.~\eqref{eq:main} does not exist for an arbitrary environment-reference state $\rho$. However, this structure simplifies under certain conditions:
\begin{corollary}
\label{cor:block_unitary}
For a code satisfying only the orthogonality condition of the KL conditions, i.e.,
\begin{equation}
   \braket{i|E_k^\dagger E_l|j} = 0 \quad (\text{for } i \neq j),
\end{equation}
the optimal disentangling unitary $U$ admits a block-diagonal form $U = \bigoplus_{i=1}^{d_L} U_i$. Because the environment-reference state decomposes into a block-diagonal form $\rho = \bigoplus_{i=1}^{d_L} \rho_i$, finding the optimal recovery map formulates as a generalized Procrustes analysis (GPA) problem, which lacks a generic closed-form solution but can be solved efficiently via numerical optimization~\cite{Gower1975,TenBerge1977}.
\end{corollary}
Based on the block-diagonal form $U$, we can derive lower bounds by strategically choosing $U$. Setting $U=I$ reduces our framework to the near-optimal entanglement fidelity constructed via the transpose map in Ref.~\cite{Near-optimal}. 
Beyond this choice, a tighter bound can be achieved by exploiting the fine structure of environment-reference state $\rho$.
Under the AQEC regime, $\rho \approx \sigma_E \otimes I_{S'}/d_L$, the dominant contribution to the entanglement fidelity arises from the diagonal blocks $\braket{i|E_k^\dagger E_l|i}$ of $\rho$. By aligning the relative phases of these diagonal blocks to induce constructive interference, we establish the following in-phase bound:
\begin{proposition}[In-phase lower bound of entanglement fidelity]
Let $\rho'_{ij}= \braket{i|\sqrt{\rho}|j}_{S'}$, $\max_\cR F_e$ is lower bounded by
\begin{align}
    F_{\text{IP}} = \frac{1}{d_L}\left[\sum_{i=1}^{d_L} \lVert\rho'_{ii}\rVert_F^2+2\sum_{i<j}\Tr(|\rho'_{ii}||\rho'_{jj}|)\right].
\end{align}
This bound is achieved by a block-diagonal unitary $U = \bigoplus_{i=1}^{d_L} U_i$, such that $U_i \rho'_{ii} = |\rho'_{ii}|$. For a single qubit code, the lower bound can be further tightened to
\begin{align}
    F_{\text{IP,qubit}} = \frac{1}{2}\left[\lVert\rho'_{11}\rVert_F^2+\lVert\rho'_{22}\rVert_F^2+2\Tr(|\rho^{\prime \dagger}_{11}\rho_{22}^{\prime}|)\right],
\end{align}
which is achieved by taking the singular value decomposition of $\rho_{11}^{\prime \dagger}\rho_{22}'$ (see Appendix~\ref{app:in-phase} for calculations).
\end{proposition}
Remarkably, for the qubit case ($d_L=2$) satisfying $\braket{0|E_k^\dagger E_l|1}=\boldsymbol{0}$, the proposed bound becomes strictly exact. 
\begin{figure}[t]
    \centering
    \includegraphics[width=1.0\linewidth]{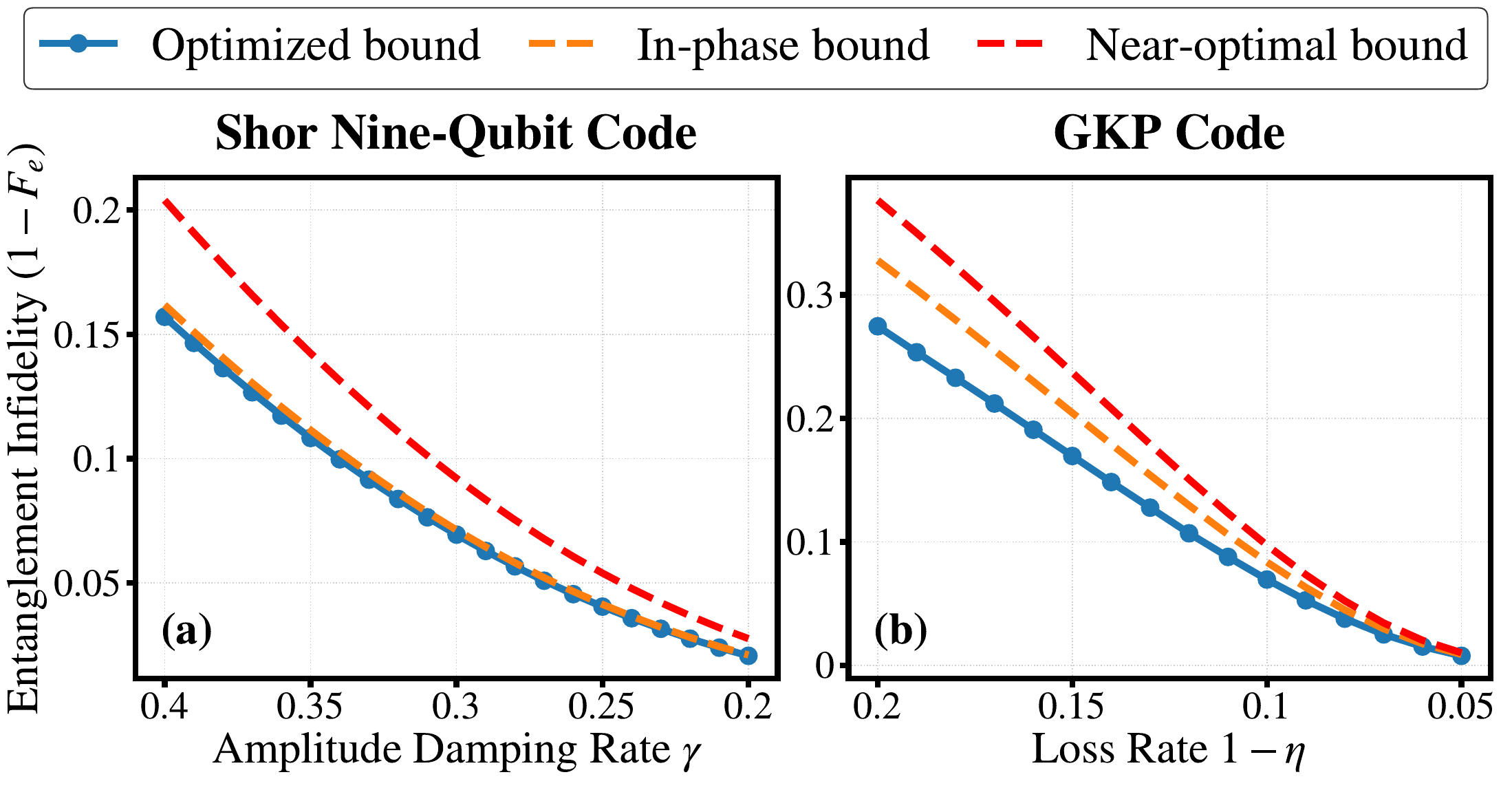}
    \caption{Comparison of entanglement infidelity lower bounds. 
    (a) Shor nine-qubit code under the amplitude damping channel as a function of the damping rate $\gamma$. 
    (b) GKP code with mean photon number $\bar{n}_{\text{GKP}}=10$ in a thermal loss channel. 
    The GKP code performance is evaluated against the channel transmissivity $\eta$ with a thermal background of $\bar{n}_{\text{th}}=1$. 
    \vspace{-5 mm}}
    \label{fig:LowerBound}
\end{figure}
To validate the proposed framework, we numerically investigate the in-phase lower bound and compare it with both the near-optimal bound attained by the transpose channel \cite{Near-optimal, Petz1986Recovery} and the fidelity by optimizing the unitary in Eq.~\eqref{eq:main}. As illustrated in Fig.~\ref{fig:LowerBound}, our benchmarks include the square-lattice GKP code subject to a thermal loss channel and the Shor nine-qubit code undergoing independent amplitude damping, respectively.
The explicit expressions of channel's Kraus operators $\{E_k\}_{k=1}^{d_E}$ and the logical code-space configurations $\{|i\rangle_S\}$ are detailed in Appendix~\ref{app:thermal-loss-amplitude-damping}. 
Across most parameter ranges of loss rate $1-\eta$ and damping rate $\gamma$, the optimal infidelity exceeds $0.01$, confining the code strictly to the AQEC regime rather than exact correction.
In both Figs.~\ref{fig:LowerBound}(a) and \ref{fig:LowerBound}(b), the near-optimal bound noticeably deviate from the numerically optimized entanglement infidelity. In contrast, the in-phase bound consistently performs better, remaining remarkably tight for the Shor nine-qubit code. In this case, the precision arises since the off-diagonal blocks of $\rho$ vanish, i.e., $\braket{i| E_k^\dagger E_l |j} \rightarrow 0$ for $i \neq j$. The reduced efficacy of the transpose channel stems from its neglect of the relative phases between the major diagonal blocks, which our in-phase alignment restores. 
For the GKP code, the in-phase lower bound begins to deviate from the exact numerically optimized fidelity in the deep thermal-loss regime where $1-\eta > 0.1$. Nevertheless, the optimization bottleneck of identifying the optimal disentangling unitary $U$ can be substantially alleviated; as detailed below, our PCA-based framework drastically accelerates the search for the optimal $U$.

\section{PCA-based framework}
To circumvent the high-dimensional complexity of the optimization, we first exploit the underlying gauge symmetry of the recovery problem. Since applying a local environmental unitary $U_E \otimes I_{S'}$ leaves the entanglement fidelity invariant, the search space for the disentangling unitary $U$ naturally reduces to the quotient manifold $\mathcal{U}(d_E d_L) / \mathcal{U}(d_E)$. Via the Cayley transform, this quotient manifold can be fully parameterized by a free anti-Hermitian matrix, requiring $d_E^2(d_L^2-1)$ real parameters. We employ this framework to determine the exact performance bounds for the Shor nine-qubit code in Fig.~\ref{fig:LowerBound}(a). However, such an exact parameterization suffers from a severe curse of dimensionality. For the GKP code with $d_E = 2415$, the optimization is highly bottlenecked yet remains feasible, whereas for marginally larger dimensions, it quickly becomes fundamentally out of reach. To resolve this challenge, we leverage the crucial observation that the optimization space can be compressed via PCA when the spectrum of the density matrix $\rho$ is highly concentrated. This PCA-based framework is formalized below:
\begin{proposition}[Fidelity bounds under PCA truncation]
\label{thm:PCA-bound}
Let $\rho = U_\rho \Sigma U_\rho^\dagger$ be the spectral decomposition of the density matrix, where $\Sigma=\Sigma_0\oplus \Sigma_1$, with $\Sigma_0$ and $\Sigma_1$ being non-negative diagonal matrices. Let $W$ denote the matrix formed by the first $M$ eigenvectors of $\rho$ corresponding to $\Sigma_0$, and let $\epsilon = \operatorname{Tr}(\Sigma_1)$ represent the trace residual. Define the compressed operator $Y = \Sigma_0^{1/2} W^\dagger$. The target isometry $V$ (satisfying $V^\dagger V = I_M$) in the PCA-based framework is given by
\begin{align}
    V = \underset{V}{\operatorname{argmax}} \left\| \operatorname{Tr}_{S'}(V Y) \right\|_F^2.
\end{align}
By extending $V$ to a full unitary matrix $U = (V \; V_\perp)$, the estimated entanglement fidelity is given by
\begin{align}
    \Tilde{F}_e = \frac{1}{d_L} \left\| \operatorname{Tr}_{S'} \left(U \Sigma^{1/2} U_\rho^\dagger \right) \right\|_F^2.
\end{align}
Then, the deviation of $\Tilde{F}_e$ from the optimal entanglement fidelity satisfies
\begin{align}
    0\leq\max_\cR F_e - \Tilde{F}_e \leq \epsilon + 4\sqrt{\epsilon(1-\epsilon)}.
    \label{eq:PCA-bound}
\end{align}
\end{proposition}
The derivation of the bound follows from the Kadison-Schwarz and Cauchy-Schwarz inequalities. A comprehensive proof is provided in Appendix~\ref{app:PCA}. For $\epsilon \ll 1$, the truncation error is well-approximated by $4\sqrt{\epsilon}$. By mapping the optimization to a lower-dimensional subspace, the PCA-based framework reduces the matrix complexity from $\mathcal{O}(d_E d_L)$ to $\mathcal{O}(M d_L)$.  To rigorously establish the parameterization, we exploit the gauge freedom: left-multiplying the target isometry $V$ by $U_E\otimes I_{S'}$ leaves the entanglement fidelity invariant. By partition-reshaping $V\in \mathbb{C}^{d_E d_L \times M}$ horizontally into a wide matrix $V' \in \mathbb{C}^{d_E \times M d_L }$, this gauge action simplifies to a linear transformation $U_E V'$. Consequently, the environmental degrees of freedom can be modded out via a polar decomposition $V' = \Lambda U'$, where $U' \in \mathcal{U}(M d_L)$. To satisfy the global isometric constraint $V^\dagger V = I_M$, the unconstrained candidate $V'$ is then mapped back to the target Stiefel manifold via a standard QR-based retraction onto the Stiefel manifold. The number of free real parameters in our parameterization scales as $(Md_L)^2+\min(Md_L,d_E)$.

\begin{figure}[t]
    \centering
    \includegraphics[width=1.0\linewidth]{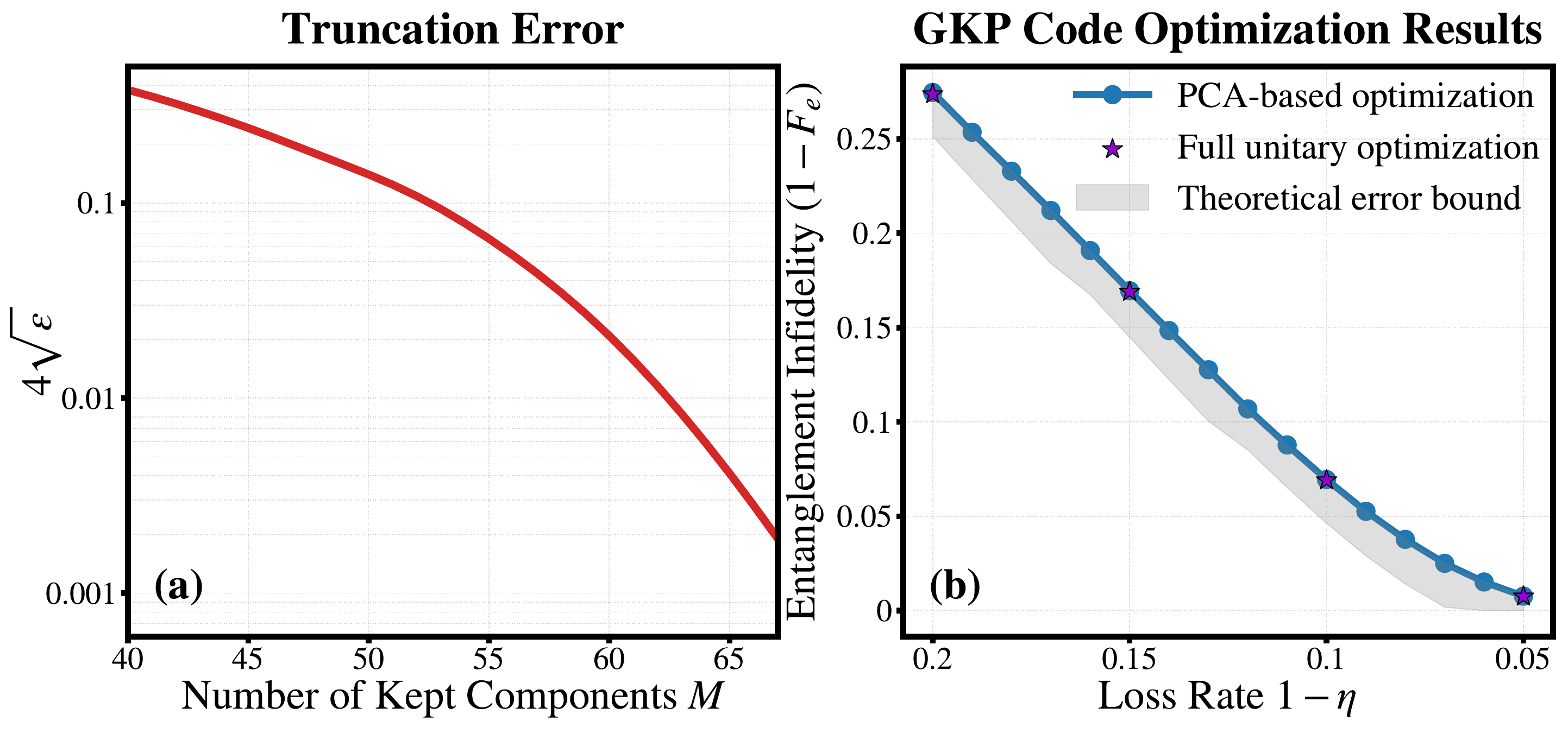}
    \caption{PCA-based optimization of entanglement infidelity. (a) Truncation error $4\sqrt{\epsilon}$ as a function of the number of principal components $M$ ($d_E=2415$) for GKP code subject to a thermal loss channel with transmission $\eta=0.87$ and $\bar{n}_{\text{th}}=1$. (b) Entanglement infidelity obtained via the PCA-based optimization compared with full unitary optimization, evaluated in thermal loss channel with $\bar{n}_{\text{th}}=1$.}
    \vspace{-5 mm}
    \label{fig:PCA}
\end{figure}

For amplitude damping channels with $\gamma=0.3$ and $N=9$, the eigenvalue spectrum of the corresponding $\rho$ remains relatively broad, with no significant separation between dominant and subdominant components. For weaker damping $\gamma < 0.1$, however, the spectrum develops a separation between the leading eigenvalues and the remaining modes, a behavior we expect to scale robustly with larger system sizes $N$. Physically, this spectral concentration arises because the weights of the Kraus operators for an $N$-qubit amplitude damping channel scale as $(\gamma/2)^{n}(1-\gamma/2)^{N-n}$, where $n\le N$ denotes the number of affected qubits (see Appendix~\ref{app:Spectrum-concentration} for scaling arguments and spectrum concentrations). For thermal loss channels, the spectral distribution of $\rho$ for GKP codes features an exponential concentration for $\bar{n}_{\text{th}}=1$. This behavior stems from the exponential decaying weights of the Kraus operators, determined by $\left[\bar{n}_{\text{th}} / (\bar{n}_{\text{th}}+1)\right]^{n}$, where $n$ is the environmental Fock-state occupation and $\bar{n}_{\text{th}}$ is the mean thermal photon number.
As illustrated in Fig.~\ref{fig:PCA}(a), this geometric scaling manifests as an exponential suppression of the bound $4\sqrt{\epsilon}$ as the number of principal components $M$ increases. Although the full environment spans a vast dimension ($d_E=2415$), the truncation error effectively vanishes within a tiny fraction of the total support ($M\ll d_E)$.

Underpinned by this concentration, PCA can be deployed to efficiently compress the optimization space. As illustrated in Fig.~\ref{fig:PCA}(b), we obtain the optimized entanglement infidelity of the GKP code. The number of principal components $M$ is dynamically adjusted between 60 and 65 across different loss rates $1-\eta$ to guarantee that the truncation error $4\sqrt{\epsilon} \le 10^{-2}$. The entanglement infidelities obtained via the PCA-based framework and the full unitary optimization exhibit excellent agreement with each other. The gray shaded region denotes the theoretical error bounds given by Eq.~\eqref{eq:PCA-bound}. In terms of computational efficiency, the isometry optimization is much faster than the unitary optimization. For both optimization schemes, we employ the Adam optimizer with a convergence tolerance of $10^{-7}$. The net elapsed CPU times for the results in Fig.~\ref{fig:PCA}(b) are benchmarked. The full unitary optimization requires an average of $2044$ seconds ($\approx 34$ minutes), whereas the proposed isometry optimization reduces the average execution time to merely $61.66$ seconds. Consequently, we demonstrate that for the thermal loss channel, our framework achieves a parameter reduction from $d_E=2415$ to $M=65$ alongside a 33-fold speedup, without sacrificing optimization accuracy.

\textit{Discussion.}---\
We have introduced a dual representation that yields, via the in-phase bound, a tighter and more broadly applicable alternative to the
transpose-channel bound, and a PCA-based framework that renders
high-dimensional AQEC optimization tractable. The in-phase bound remains
tight for the Shor nine-qubit code under amplitude damping. For systems with $d_L\geq3$, this bound can be tightened further via GPA. 

The effectiveness of the PCA framework is governed by the spectral concentration of $\rho$, determined by the decay rate of the channel's Kraus operators. For multi-qubit DV codes, the relevant decay is determined by the number of affected qubits: $(\gamma/2)^{n}(1-\gamma/2)^{N-n}$. We observe that the eigenvalue spectrum becomes concentrated in the weak-damping regime ($\gamma< 0.1$) and expect such concentration to emerge in the large-$N$ regime, $N \gg 1/\vert{}\ln(1-\gamma)\vert{}$. 
For the GKP code under thermal loss, the decaying weights of Kraus operators, $[\bar{n}_{\rm th}/(\bar{n}_{\rm th}+1)]^{n}$, lead to a concentrated eigenvalue spectrum and a reported 33-fold numerical speedup. This mechanism naturally extends to other CV models with exponentially decaying weights of Kraus operators, including small-displacement, photon-loss, and pure-dephasing channels. More broadly, the exponential concentration of the error spectrum recurs whenever a code couples to a thermal-like environment: such as the Gibbs occupancy $e^{-\beta E_n}$ in thermal baths, or the Boltzmann-suppressed anyonic excitations $e^{-k\beta\Delta}$ in topological phases. Within the AdS/CFT correspondence, an analogous concentration arises from the low-energy effective-field-theory cutoff on bulk operators. These connections suggest that our isometry optimization on quotient manifolds may serve as a universal computational tool for AQEC across diverse physical regimes beyond quantum computing.

\textit{Acknowledgment.}---\
 This work is supported by the Fermi Forward Discovery Group LLC under Contract No. FWP-23-24 with the U.S. Department of Energy, Office of Science, Advanced Scientific Computing Research (ASCR) Program. S.Z. acknowledges support by the DOE's Early Career Research Program and by the U.S. Department of Energy. D.K. acknowledges support by the U.S. Department of Energy, Office of Science, National Quantum Information Science Research Centers, Superconducting Quantum Materials and Systems Center (SQMS) under Contract No. 89243024CSC000002, M.G. is supported by CERN through the CERN Quantum Technology Initiative. 
 
 \textit{Data Availability.---}The optimization source code and data-generation scripts used in this work are publicly available at https://github.com/JingWu-31/QEC-fidelity under the MIT license.

\bibliographystyle{quantum}
\bibliography{ref}

\onecolumn
\appendix

\maketitle

\section{Proof of the dual representation}
\label{app1}
We consider the composite system $\mathcal{H}_S \otimes \mathcal{H}_{S'} \otimes \mathcal{H}_A \otimes \mathcal{H}_E$. Here, $S$ denotes the encoded system, $S'$ is the reference system forming a purified pair with $S$, $E$ represents the environment degrees of freedom, and $A$ is the ancilla system utilized by the recovery map $\mathcal{R}$. 
By Uhlmann theorem [19], the $\max_{\sigma_E}{\sqrt{F}}$ in Eq. (4) is given by
\begin{align}
    \frac{1}{\sqrt{d_L}} \max_{\sigma_E} \max_{U'}    |\Tr_{ES'} [U'\sqrt{\rho}(\sqrt{\sigma_E}\otimes I_{S'})] |, \label{eq:double-max}
\end{align}
where $U'$ is the optimal unitary on $ES'$ in Uhlmann theorem.
Using the trace identity $\Tr_{AB}[X_{AB} (Y_A \otimes I_B)] = \Tr_A[\Tr_B(X_{AB})Y_A]$ and defining $X'=\Tr_{S'}(U'\sqrt{\rho})$, we obtain
\begin{align}
    \max_{\sigma_E}{\sqrt{F}}=\frac{1}{\sqrt{d_L}} \max_{U'} \max_{\sigma_E} |\Tr_E (X'\sqrt{\sigma_E})|.
    \label{eq:S1-maxF1}
\end{align}
The unitary $U'$ possesses sufficient degrees of freedom to accommodate an environmental unitary $U_E$. Let $U' = (U_E\otimes I_{S'})U$, then
\begin{align}
    X' &= \Tr_{S'}[(U_E \otimes I_{S'})U\sqrt{\rho}] = U_E \Tr_{S'}(U\sqrt{\rho}).
    \label{eq:app:S1-2}
\end{align}
Let $X=\Tr_{S'}(U\sqrt{\rho})$, from Eqs. \eqref{eq:S1-maxF1} and \eqref{eq:app:S1-2}, we have
\begin{align}
    \max_{\sigma_E}{\sqrt{F}}=\frac{1}{\sqrt{d_L}} \max_{U}\{ \max_{U_E}\max_{\sigma_E}|\Tr_E U_EX\sqrt{\sigma_E}|\}.
    \label{eq:app1-1}
\end{align}
Inside the bracket,
\begin{align}
    \max_{U_E}\max_{\sigma_E}|\Tr_E U_EX\sqrt{\sigma_E}| = \| X\|_F,
    \label{eq:app1-2}
\end{align}
where $\| \cdot\|_F$ is the Frobenius norm. This equality can be shown by Von Neumann's trace inequality: 
\begin{align}
|\Tr_E U_EX\sqrt{\sigma_E}|\leq \sum_i \alpha_i \beta_i,
\end{align}
where $\alpha_i$ and $\beta_i$ are the singular values of the matrices $U_E X$ and $\sqrt{\sigma_E}$, respectively. The equality holds if and only if the two matrices are aligned:
\begin{align}
    U_E X = W\Lambda_\alpha W^\dagger, \quad \sqrt{\sigma_E} = W \Lambda_\beta W^\dagger, 
\end{align}
and $\beta_i = \alpha_i/\sqrt{\sum_j \alpha_j^2}$. Therefore, the maximum can be obtained by
\begin{align}
    U_E = |X|X^+, \quad \sigma_E = |X|^2/ \Tr(|X|^2), 
    \label{eq:app-optimal-sigma-U_E}
\end{align}
where $(\cdot)^+$ denotes the Moore-Penrose pseudoinverse. Combining Eq. \eqref{eq:app1-1} and Eq. \eqref{eq:app1-2}, we obtain
\begin{align}
    \max_{\sigma_E}{\sqrt{F}} =\frac{1}{\sqrt{d_L}}\max_U\|\Tr_{S'}(U\sqrt{\rho})\|_F.
    \label{eq:app-main}
\end{align}
Once the unitary $U$ in Eq. \eqref{eq:app-main} is obtained, the corresponding recovery map $\mathcal{R}$ is uniquely determined. Formally, $\mathcal{R}$ is implemented by a unitary operator $U_{\mathcal{R}}$ acting on the joint Hilbert space of the system $S$ and ancilla $A$ given by:
\begin{equation}
\mathcal{R}(\rho_S) = \Tr_A \left[ U_{\mathcal{R}} (\rho_S \otimes \ketbra{0}{0}_A) U_{\mathcal{R}}^\dagger \right].
\end{equation}
According to the duality theorem in [28],
\begin{align}
    \max_{\cR} F(\cR\circ\cN,\cI) = \max_{\cR'} F(\cN^C,\cR' \circ\cI^C),
\end{align}
which holds for any fixed purification of the input state $\ket{\psi}_{SS'}$. In particular, the $F_e$ is designated for the case where the input state is the maximally entangled state.
The optimal recovery unitary $U_{\mathcal{R}}$ acting on the joint system $SA$ corresponds precisely to the Uhlmann unitary obtained via purification, which is associated with the quantum fidelity $F_e(\mathcal{N}^C, \mathcal{R}'\circ\cI^C)$.
To explicitly identify $U_{\mathcal{R}}$, we invoke the fidelity identity derived from Uhlmann's theorem:
\begin{align}
    \sqrt{F}&=\max_{U_C} |\braket{\psi_{\rho_B} |I_B \otimes U_C |\psi_{\sigma_B}}| \nonumber \\
    &= \max_{U_B}|\Tr_B(U_B \sqrt{\rho_B}\sqrt{\sigma_B})|,
    \label{eq:app-2}
\end{align}
where the states $\rho_B$ and $\sigma_B$ are defined on system $B$, and the system $C$ denotes the reference system that purifies $\rho_B$ and $\sigma_B$ into $\ket{\psi_{\rho_B}}$ and $\ket{\psi_{\sigma_B}}$, respectively. 
In the above identity,  $U_B=U_C^\T$. 
Identifying the respective sub-systems as $B = ES'$ and $C = SA$, a direct comparison between Eq.~\eqref{eq:double-max} and Eq.~\eqref{eq:app-2} yields
\begin{align}
    U_\cR^\T = U'=(U_E\otimes I_{S'})U.
\end{align}
To ensure a well-defined Stinespring dilation for the recovery map $\mathcal{R}$, the dimension of the ancilla space $A$ must be large enough to accommodate the full matrix rank of the recovery unitary $U_{\mathcal{R}}$.
From Eq. \eqref{eq:app-optimal-sigma-U_E}, we finally obtain
\begin{align}
    U_\cR^\T = (|X|X^+\otimes I_{S'}) U.
\end{align}

\section{Calculations of the in-phase lower bound}
\label{app:in-phase}
In the basis of $S'$, the matrix $\rho'_{ij} = \langle i \vert{} \sqrt{\rho} \vert{} j \rangle$ exhibits a block structure:
\begin{align}
    \rho' = \begin{pmatrix}
        \rho'_{11} & \rho'_{12} & \cdots \\
        \rho'_{21} & \rho'_{22} & \cdots \\
        \vdots & \vdots & \ddots
    \end{pmatrix}.
\end{align}
Heuristically, for an effective quantum error-correcting code, the off-diagonal blocks $\rho'_{i \neq j} \to 0$. Given a block-diagonal unitary $U = \bigoplus_{i} U_i$, we have
\begin{align}
    \text{Tr}_{S'}(U\sqrt{\rho}) = \sum_{i=1}^{d_L} U_i \rho'_{ii}.
\end{align}
Consequently, the squared Frobenius norm of this partial trace expands as
\begin{align}
    \left\Vert \sum_i U_i \rho'_{ii} \right\Vert_F^2 = \sum_i \left\Vert \rho'_{ii} \right\Vert_F^2 + \sum_{i \neq j} \text{Tr}\big( \rho_{ii}^{\prime\dagger} U_i^\dagger U_j \rho'_{jj} \big).
\end{align}
By employing the polar decomposition $U_i \rho'_{ii} = \vert{}\rho'_{ii}\vert{}$, the cross-terms simplify as
\begin{align}
    \text{Tr}\big(\rho_{ii}^{\prime\dagger} U_i^\dagger U_j \rho'_{jj} \big) = \text{Tr}\big( |\rho'_{ii}| |\rho'_{jj}| \big).
\end{align}
Consequently, we establish the in-phase lower bound on the entanglement fidelity:
\begin{align}
    F_e &= \max_U \frac{1}{d_L} \left\Vert \text{Tr}_{S'}(U\sqrt{\rho}) \right\Vert_F^2 \\
    &\geq \frac{1}{d_L} \left[ \sum_i \left\Vert \rho'_{ii} \right\Vert_F^2 + 2\sum_{i \neq j} \text{Tr}\big( |\rho'_{ii}| |\rho'_{jj}| \big) \right].
\end{align}
Specifically, for a qubit, the squared Frobenius norm expands as
\begin{align}
    \left\Vert \rho'_{11} \right\Vert_F^2 + \left\Vert \rho'_{22} \right\Vert_F^2 + 2\Re \text{Tr}\left( \rho_{11}^{\prime\dagger} U_1^\dagger U_2 \rho'_{22} \right).
\end{align}
The cross-term can be maximized over the unitaries $U_1$ and $U_2$ as
\begin{align}
    \max_{U_1,U_2}2\Re \text{Tr}\left( \rho_{11}^{\prime\dagger} U_1^\dagger U_2 \rho'_{22} \right) = 2\Tr\left(|\rho_{11}^{\prime \dagger}\rho_{22}'|\right),
\end{align}
which is achieved by setting $U_1 = I$ and choosing $U_2$ from the polar decomposition of the operator product, i.e., $U_2 \rho'_{22}\rho_{11}^{\prime \dagger} = \vert{}\rho'_{22}\rho_{11}^{\prime \dagger}\vert{}$. Therefore we obtain the lower bound for qubit
\begin{align}
    F_{\text{IP,qubit}} = \frac{1}{2}\left[\lVert\rho'_{11}\rVert_F^2+\lVert\rho'_{22}\rVert_F^2+2\Tr(|\rho^{\prime \dagger}_{11}\rho_{22}^{\prime}|)\right].
\end{align}
\vspace{5 mm }
\section{Thermal loss channel and amplitude damping channel}
\label{app:thermal-loss-amplitude-damping}
\subsection{Thermal loss channel}
A thermal loss channel $\mathcal{N}$ is characterized by the transmissivity $\eta = \cos^2\theta$ and the mean photon number $\bar{n}_{\text{th}}$ of the environment. Let $\hat{a}$ and $\hat{b}$ be the annihilation operators for the bosonic system $A$ and the environment $B$, respectively. The channel is defined as:
\begin{align}
\mathcal{N}(\rho) = \text{Tr}_B \left( \hat{U} (\rho \otimes \rho_{\text{th}}) \hat{U}^\dagger \right),
\label{eq:app-thermal-loss-definition}
\end{align}
where the initial state of the environment is a thermal state:
\begin{align}
\rho_{\text{th}} = \sum_{n=0}^{\infty} \frac{\bar{n}_{\text{th}}^n}{(\bar{n}_{\text{th}}+1)^{n+1}} |n\rangle\langle n|_B,
\label{eq:app-thermal-state}
\end{align}
and the beam-splitter interaction $\hat{U}$ is given by:
\begin{align}
\hat{U} = \exp\left[ i\theta (\hat{a}^\dagger \hat{b} + \hat{a} \hat{b}^\dagger) \right].
\end{align}
The Kraus operators $\{E_{m,n}\}$ of the thermal loss channel can be derived from Eq. \eqref{eq:app-thermal-loss-definition} by projecting the joint unitary onto the environment's initial thermal basis $|n\rangle_B$ and final basis $|m\rangle_B$:
\begin{align}
    E_{m,n}  = \sqrt{\frac{\bar{n}_{\text{th}}^n}{(\bar{n}_{\text{th}}+1)^{n+1}}} \braket{m|\hat{U}|n}_B.
    \label{eq:app-S2-Emn}
\end{align}
Let $U_{m,n} = \braket{m|\hat{U}|n}$. Using the disentanglement theorem for the $SU(2)$ Lie algebra,
\begin{align}
    &U = e^{(i \tan \theta) a^\dagger b} (\sec \theta)^{a^\dagger a - b^\dagger b} e^{(i \tan \theta) b^\dagger a},\\
   & U = e^{(i \tan \theta) b^\dagger a} (\cos \theta)^{a^\dagger a - b^\dagger b} e^{(i \tan \theta) a^\dagger b},
\end{align}
the matrix elements of $U_{m,n}$ are given by Jacobi polynomial $P_n^{(\alpha,\beta)}$

\begin{align}
    \braket{k,m|U|j,n} = \delta_{j+n,k+m} \sqrt{\frac{n!j!}{k!m!}} \left(i \sqrt{\frac{1-\eta}{\eta}}\right)^{m-n}\eta^{\frac{j-n}{2}}P_n^{(m-n,j-m)}(2\eta-1).
    \label{eq:app-S2-Uinitial}
\end{align}

For efficient numerical implementation, particularly when dealing with large Fock space truncations, the matrix elements $U_{m,n} = \langle m| \hat{U} |n\rangle_B$ are generated using a recursive relation. By applying the Heisenberg evolution of the annihilation operator $\hat{U}^\dagger \hat{b} \hat{U} = i\sin\theta \hat{a} + \cos\theta \hat{b}$, we derive the following recurrence formula:
\begin{align}
    U_{m+1,n} &= \bra{m+1}U\ket{n}=\frac{1}{\sqrt{m+1}}\bra{m} \hat{b} \hat{U} \ket{n} \nonumber \\
    & =\frac{1}{\sqrt{m+1}}\bra{m}\hat{U}((i\sin\theta \hat{a}+\cos\theta \hat{b})\ket{n} \nonumber \\
    &=\frac{i\sin\theta}{\sqrt{m+1}} U_{m,n} \hat{a}+\frac{\cos \theta \sqrt{n}}{\sqrt{m+1}} U_{m,n-1}.
    \label{eq:app-S2-Urecurrence}
\end{align}
The Kraus operators of the thermal loss channel are generated using Eqs.~\eqref{eq:app-S2-Emn} and \eqref{eq:app-S2-Urecurrence}, with the recurrence relation initialized by $U_{m,n}$ from Eq.~\eqref{eq:app-S2-Uinitial}.

The beam-splitter interaction $\hat{U}$ preserves the total photon number of the system and environment. For a numerical simulation involving a system Hilbert space truncated at $N_S$ photons and an environment truncated at $N_E$ photons, the total photon number is bounded by $N_{\text{tot}} = N_S + N_E$. Consequently, we must account for all photon switching processes $\hat{U}$ where the initial and final environment states $(n, m)$ satisfy $m + n \leq N_{\text{tot}}$. The total number of Kraus operators required to fully describe the channel dynamics scales approximately as 
\begin{align}
    d_E \approx (N_S + N_E)^2.
    \label{eq:app-d_E}
\end{align}
In the thermal loss channel $\bar{n}_{\text{th}}=1$, we take $N_E = 9$ so that the tail probability of the thermal state in Eq. \eqref{eq:app-thermal-state} is less than $0.2\%$.
The Wigner function of square-lattice GKP qubit code is given by 
\begin{equation}
\psi_{\mu}(x) = N_\mu \sum_{s \in \mathbb{Z}} e^{-\frac{1}{2}\kappa^2 \bar{x}_s^2} e^{-\frac{(x - \bar{x}_s)^2}{2\Delta^2}},
\end{equation}
where $\mu \in\{0,1\}$, $N_\mu$ is the normalization factor, $\bar{x}_s = (2s + \mu)\sqrt{\pi}$ and $\kappa=\Delta$ defines the energy of the GKP code. In our numerical simulations, we project the $\psi_{\mu}$ into the Fock basis. We select $\kappa=\Delta=0.2159$ for $\bar{n}_{\text{GKP}}=10$ and $N_S=60$, so that the tail probability of the state is strictly bounded: 
\begin{equation}
\sum_{n=N_S+1}^{\infty} P(n) \leq 0.5\%,
\end{equation}
where $P(n) = |\langle n | \psi \rangle|^2$.
\subsection{Amplitude damping channel}
The amplitude damping channel $\mathcal{E}$ is characterized by the decay probability $\gamma \in [0, 1]$. For a single qubit, the channel is described by the Kraus operators:
\begin{align}
E_0 = \begin{pmatrix} 1 & 0 \\ 0 & \sqrt{1-\gamma} \end{pmatrix}, \quad
E_1 = \begin{pmatrix} 0 & \sqrt{\gamma} \\ 0 & 0 \end{pmatrix}.
\label{eq:app-AmplitudeDamping1}
\end{align}
For the Shor nine-qubit code, we assume that each physical qubit undergoes independent and identical amplitude damping. The resulting multi-qubit Kraus operators $\mathbf{E}_{\mathbf{k}}$ are given by the tensor product of the individual Kraus operators:
\begin{align}
E_{\mathbf{k}} = E_{k_1}^{(1)} \otimes E_{k_2}^{(2)} \otimes \dots \otimes E_{k_9}^{(9)},
\label{eq:app-AmplitudeDamping2}
\end{align}
where the vector index $\mathbf{k} = (k_1, k_2, \dots, k_9)$ spans the set $\{0, 1\}^9$. The total number of the Kraus operators is given by
\begin{equation}
    d_E = 2^9 = 512.
\end{equation}
The logical qubit states of the Shor nine-qubit code are given by
\begin{align}
\ket{\Bar{0}} &= \frac{1}{\sqrt{8}} (|000\rangle + |111\rangle)(|000\rangle + |111\rangle)(|000\rangle + |111\rangle), \\
\ket{\Bar{1}} &= \frac{1}{\sqrt{8}} (|000\rangle - |111\rangle)(|000\rangle - |111\rangle)(|000\rangle - |111\rangle).
\end{align}

\section{Proof of Fidelity bounds under PCA truncation}
\label{app:PCA}
In the PCA-based framework, we partition the underlying Hilbert space dimension to its principal and residual components. Here, the total dimension of the composite space is $N = d_E d_L$, and $M \le d_E$ denotes the number of principal components chosen to truncate the dominant support of the environment-reference state. Let $U \in \mathcal{U}(N)$ be an unitary operator parameterized in the block-isometry form $U = \begin{pmatrix} V & V_{\perp} \end{pmatrix}U_\rho^\dagger$, where $V \in \text{St}(M, N)$ and $V_{\perp} \in \text{St}(N-M, N)$ satisfy the standard Stiefel manifold constraints $V^\dagger V = I_M$ and $V_{\perp}^\dagger V_{\perp} = I_{N-M}$, respectively. 
The spectral decomposition of the density matrix $\rho = U_{\rho} \Sigma U_{\rho}^{\dagger}$  in the main text is partitioned such that $U_{\rho} = \begin{pmatrix} W & W_{\perp} \end{pmatrix}$ and $\Sigma = \Sigma_0 \oplus \Sigma_1$.
Substituting the explicit block structures, the reduced operator $X$ has two components:
\begin{align}
X &= \Tr_{S'}\left[ \begin{pmatrix} V & V_{\perp} 
\end{pmatrix} U_\rho^\dagger U_\rho
\begin{pmatrix} \Sigma_0^{1/2} & 0 \\ 0 & \Sigma_1^{1/2} 
\end{pmatrix} 
\begin{pmatrix} W^\dagger \\\ W_{\perp}^\dagger \end{pmatrix} \right] \nonumber\\
&= \Tr_{S'}(V \Sigma_0^{1/2} W^\dagger) + \Tr_{S'}(V_{\perp} \Sigma_1^{1/2} W_{\perp}^\dagger). \label{eq:sm_block_multiply}
\end{align}
We define
\begin{align}
X_0 &\equiv \text{Tr}_{S'}\left(V \Sigma_0^{1/2} W^\dagger\right), \\
X_1 &\equiv \text{Tr}_{S'}\left(V_{\perp} \Sigma_1^{1/2} W_{\perp}^\dagger\right).
\end{align}
By expanding the Frobenius norm via the trace identity, we obtain
\begin{align}
\| X\|_F^2 = \| X_0\|_F^2 + \| X_1\|_F^2 + 2\Re \operatorname{Tr}(X_0 X_1^\dagger). \label{eq:app-PCA-diff}
\end{align}
By Kadison-Schwarz inequality, for any operator $A$ and unital complete positive map $\phi$
\begin{align}
    \phi(A)^\dagger \phi(A) \leq \phi(A^\dagger A).
\end{align}
Taking $\phi =1/d_L\Tr_{S'}$, we obtain 
\begin{align}
 \Tr_{S'}(A)^\dagger \Tr_{S'}(A) \leq d_L \Tr_{S'}(A^\dagger A).   
\end{align}
Therefore $\Vert X_1 \Vert_F^2$ is bounded by
\begin{align}
    &X_1^\dagger X_1 = \Tr_{S'}(V_{\perp} \Sigma_1^{1/2} W_{\perp}^\dagger)^\dagger \Tr_{S'}(V_{\perp} \Sigma_1^{1/2} W_{\perp}^\dagger) \nonumber\\
    & \qquad \quad \leq d_L\Tr_{S'}(W_\perp \Sigma_1 W_\perp^\dagger), \\
    &\| X_1\|_F^2 =\Tr(X_1^\dagger X_1) \leq d_L \Tr(W_\perp \Sigma_1 W_\perp^\dagger)=d_L \epsilon.
    \label{eq:app-PCA-bound1}
\end{align}
Similarly, we have
\begin{equation}
    \|X_0\|_F^2 \leq d_L(1-\epsilon).
\end{equation}
And by Cauchy-Schwarz inequality,
\begin{align}
    2\left|\Re \Tr(X_0X_1^\dagger)\right|&\leq  2|{\Tr} (X_0X_1^\dagger)| \leq 2 \| X_0\|_F\| X_1\|_F \nonumber\\
    &\leq 2 d_L \sqrt{\epsilon(1-\epsilon)}. 
    \label{eq:app-PCA-bound2}
\end{align}
Let $U' = \begin{pmatrix} V' & V_{\perp}' \end{pmatrix}U_\rho^\dagger$ be the global unitary that maximizes the total reduced norm, denoting the maximum value as $\|X'\|_F^2$. For comparison, let $U = \begin{pmatrix} V & V_{\perp} \end{pmatrix} U_\rho^\dagger$ be the unitary whose principal block $V$ maximizes $\|X_0\|_F^2$, and let $\|X\|_F^2$ represent the corresponding full reduced norm evaluated under this local optimization choice.
The estimated entanglement fidelity by $U$ is given by
\begin{equation}
    \Tilde{F}_e = \frac{1}{d_L} \left\Vert \Tr_{S'} (U\Sigma^\frac{1}{2} U_\rho^\dagger) \right\Vert_F^2.
\end{equation}
From Eq. \eqref{eq:app-PCA-diff}, the difference between the two configurations can be expanded as 
\begin{align}
    \| X' \|_F^2 - \| X\|_F^2&= \| X'_0 \|_F^2-\| X_0 \|_F^2 + \| X'_1 \|_F^2-\| X_1 \|_F^2 \nonumber\\
    & +2\text{Re}\text{Tr}(X'_0 X_1^{\prime\dagger})-2\text{Re}\text{Tr}(X_0 X_1^\dagger).
\end{align}
By definition, $V$ maximizes the principal term, which implies $\| X'_0 \|_F^2 - \| X_0 \|_F^2 \le 0$. From Eq. \eqref{eq:app-PCA-bound1} and Eq.~\eqref{eq:app-PCA-bound2}, 
\begin{align}
&\| X'_1 \|_F^2 - \| X_1 \|_F^2 \leq d_L \epsilon,\\
&2\text{Re}\text{Tr}(X'_0 X_1^{\prime\dagger})-2\text{Re}\text{Tr}(X_0 X_1^\dagger) \leq 4d_L\sqrt{\epsilon(1-\epsilon)}.
\end{align}
Since $\max_\mathcal{R}F_e = 1/d_L \|X'\|_F^2$ and $\Tilde{F}_e = 1/d_L \|X\|_F^2$,  we divide both sides by $d_L$ to arrive at the final fidelity gap:
\begin{equation}
    \max_\mathcal{R}F_e-\Tilde{F}_e \leq \epsilon +4 \sqrt{\epsilon(1-\epsilon)}.
\end{equation}
\vspace{5 mm}
\section{Scaling of Kraus Operators and Spectral Concentration for the Nine- and Eleven-Qubit Codes}
\label{app:Spectrum-concentration}
\subsection{Scaling of Kraus operators for the thermal loss channel}
The spectrum of the environment-reference state $\rho$ can be characterized by its matrix elements $\text{Tr}(E_k \vert{}i\rangle_L \langle j\vert{} E_l^\dagger)$, where $\vert{}i\rangle_L$ denotes the basis of the logical subspace. Let $\rho_L = P_C / d_L$ be the maximally mixed state within the coding space, where $P_C = \sum_i \vert{}i\rangle_L \langle i\vert{}$ is the projector onto the code. We define the weight of each Kraus operator as the probability of the error event $E_k$ by
\begin{equation}
    p(E_k) = \text{Tr}(E_k \rho_L E_k^\dagger) = \frac{1}{d_L} \Tr(E_k^\dagger E_k P_C). 
    \label{eq:app-def-weight}
\end{equation}
For thermal loss channel, according to Eq.~\eqref{eq:app-S2-Emn}, the weight is given by
\begin{equation}
    p(E_{n,m}) = \frac{\bar{n}_{\text{th}}^n}{d_L(\bar{n}_{\text{th}}+1)^{n+1}}\Tr( A_{n,m}^\dagger A_{n,m} P_C),
\end{equation}
where $A_{n,m}=\braket{m|\hat{U}|n}_B$ is a sub-matrix of the larger beam-splitter unitary matrix $U$. Consequently, $0\leq A^\dagger_{n,m}A_{n,m}\leq 1$. Since $0\leq P_C\leq I$ is a projector, then
\begin{equation}
    \Tr(A_{n,m}^\dagger A_{n,m} P_C)\leq \Tr(A^\dagger_{n,m} A_{n,m})\leq 1.
\end{equation}
Therefore $p(E_{n,m})$ is of order $[\bar{n}_{\text{th}}/(\bar{n}_{\text{th}}+1)]^n$.
\subsection{Scaling of Kraus operators for the amplitude damping channel}
From the expressions in Eq.~\eqref{eq:app-AmplitudeDamping1}, the Kraus operators for a single-qubit amplitude damping channel satisfy
\begin{align}
    &E_0^\dagger E_0 = \left(1-\frac{\gamma}{2}\right)I + \frac{\gamma}{2}Z,\\
    &E_1^\dagger E_1 = \frac{\gamma}{2}(I-Z),
\end{align}
where $Z$ denotes the Pauli-$Z$ matrix. For an $N$-qubit system, consider an error configuration where $n$ qubits undergo the damping error $E_1$. Let $\mathcal{S}_1$ be the index set of qubits subject to $E_1$, and $\mathcal{S}_0$ be the remaining set associated with $E_0$, such that $\mathcal{S}_1 \cup \mathcal{S}_0 = \{1, 2, \dots, N\}$. The corresponding $N$-qubit Kraus weight operator expands as

\begin{align}
    E_{\boldsymbol{k}}^\dagger E_{\boldsymbol{k}} =&  \left(\bigotimes_{i \in \mathcal{S}_1} \frac{\gamma}{2}(I - Z_i)\right) \otimes \left( \bigotimes_{j \in \mathcal{S}_0} \left[ \left(1 - \frac{\gamma}{2}\right) I + \frac{\gamma}{2} Z_j \right] \right)\nonumber \\
    &= \sum_{\boldsymbol{u} \subseteq \mathcal{S}_1} \left( \frac{\gamma}{2} \right)^n (-1)^{\vert{}\boldsymbol{u}\vert{}} Z_{\boldsymbol{u}} \sum_{\boldsymbol{v} \subseteq \mathcal{S}_0}  \left( \frac{\gamma}{2} \right)^{\vert{}\boldsymbol{v}\vert{}} \left( 1 - \frac{\gamma}{2} \right)^{N - n - \vert{}\boldsymbol{v}\vert{}} Z_{ \boldsymbol{v}}\nonumber\\
    & = \left( \frac{\gamma}{2} \right)^n \left(1- \frac{\gamma}{2} \right)^{N-n}\sum_{\boldsymbol{u} \subseteq \mathcal{S}_1} \sum_{\boldsymbol{v} \subseteq \mathcal{S}_0} (-1)^{\vert{}\boldsymbol{u}\vert{}} \left(\frac{\gamma/2}{1-\gamma/2}\right)^{|\boldsymbol{v}|} Z_{\boldsymbol{u} \cup \boldsymbol{v}},
\end{align}

where $\boldsymbol{u}$ and $\boldsymbol{v}$ are subsets of $\mathcal{S}_1$ and $\mathcal{S}_0$, respectively. $Z_{\boldsymbol{u}} = \bigotimes_{k \in \boldsymbol{u}} Z_k$ represents the multi-qubit Pauli-$Z$ error supported on the subset $\boldsymbol{u}$ and $|\boldsymbol{u}|$ is the number of the elements in the subset $\boldsymbol{u}$. Since $\gamma<1$, the term $$\left(\frac{\gamma/2}{1-\gamma/2}\right)^{\vert{}\boldsymbol{v}\vert{}}$$
decreases with $\vert{}\boldsymbol{v}\vert{}$. Since Pauli operators satisfy the trace orthogonality condition, the number of non-zero terms for $\text{Tr}(Z_{\boldsymbol{u} \cup \boldsymbol{v}}P_{\mathcal{C}})$ is strictly confined to the number of pure $Z$ operators within the stabilizer group. Under the assumption of a sparse stabilizer configuration, this total summation is bounded on the order of $\mathcal{O}(1)$. Thus, for the amplitude damping channel, under the assumptions of $\gamma<1$ and a sparse stabilizer configuration, the probability of the error event $E_{\boldsymbol{k}}$ scales as
\begin{equation}
p(E_{\boldsymbol{k}})\propto \left( \frac{\gamma}{2} \right)^n \left(1- \frac{\gamma}{2} \right)^{N-n}.
\end{equation}

\begin{figure}[t]
    \centering
    \includegraphics[width=1.0\linewidth]{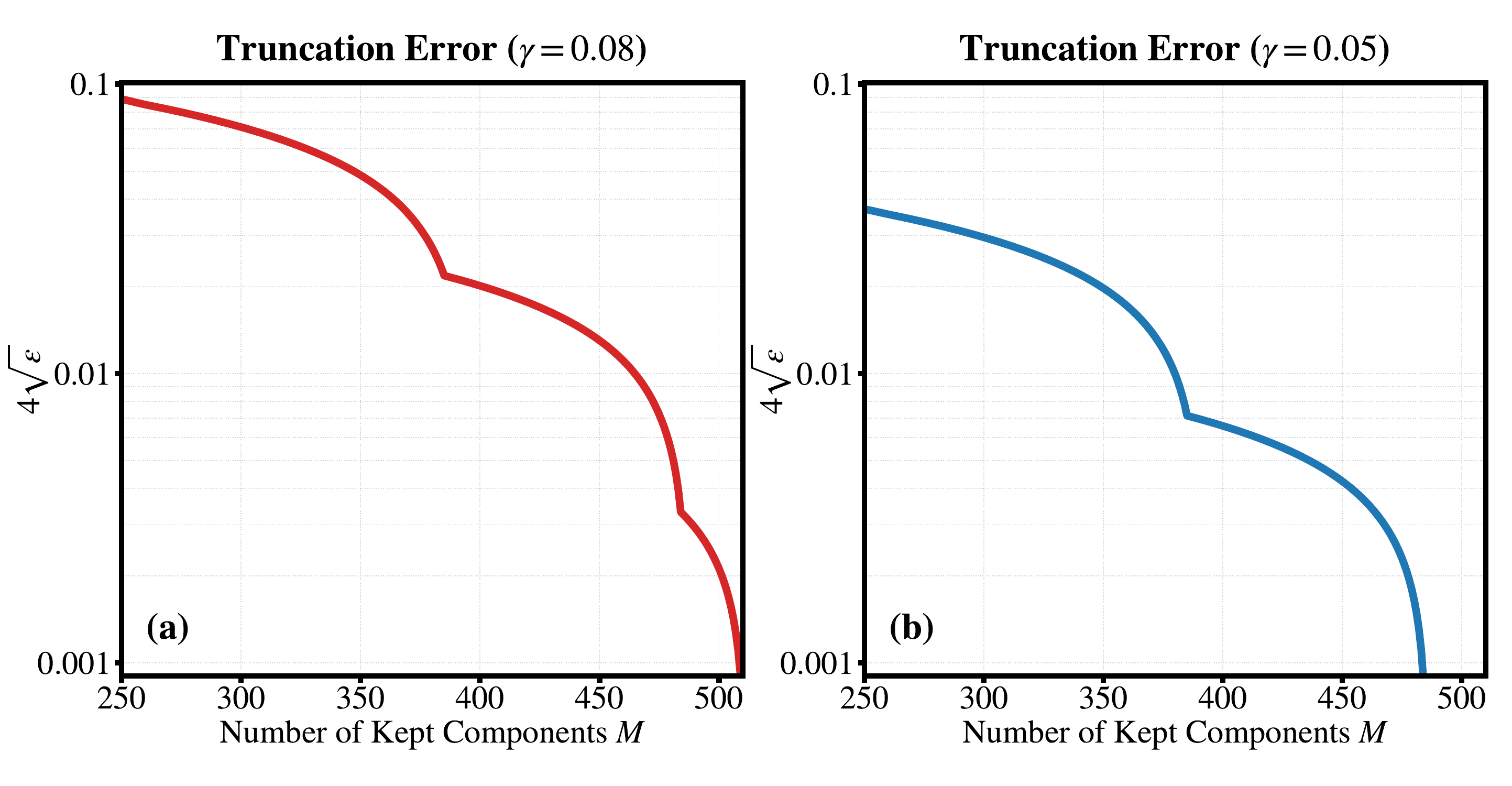}
    \caption{Truncation error $4\sqrt{\epsilon}$ as a function of the number of principal components $M$ for Shor nine-qubit code subject to amplitude damping channels (a) $\gamma = 0.08$. (b) $\gamma=0.05$.}
    \vspace{-5 mm}
    \label{fig:Truncation9}
\end{figure}

In the weak damping regime where $\gamma <0.1$, we present numerical results demonstrating the spectral concentration of the Shor nine-qubit code under the amplitude damping channel as in Fig.~\ref{fig:Truncation9}. This verifies that the spectrum is concentrated to enable the deployment of the PCA-based framework. 

\begin{figure}[tb]
    \centering
    \includegraphics[width=1.0\linewidth]{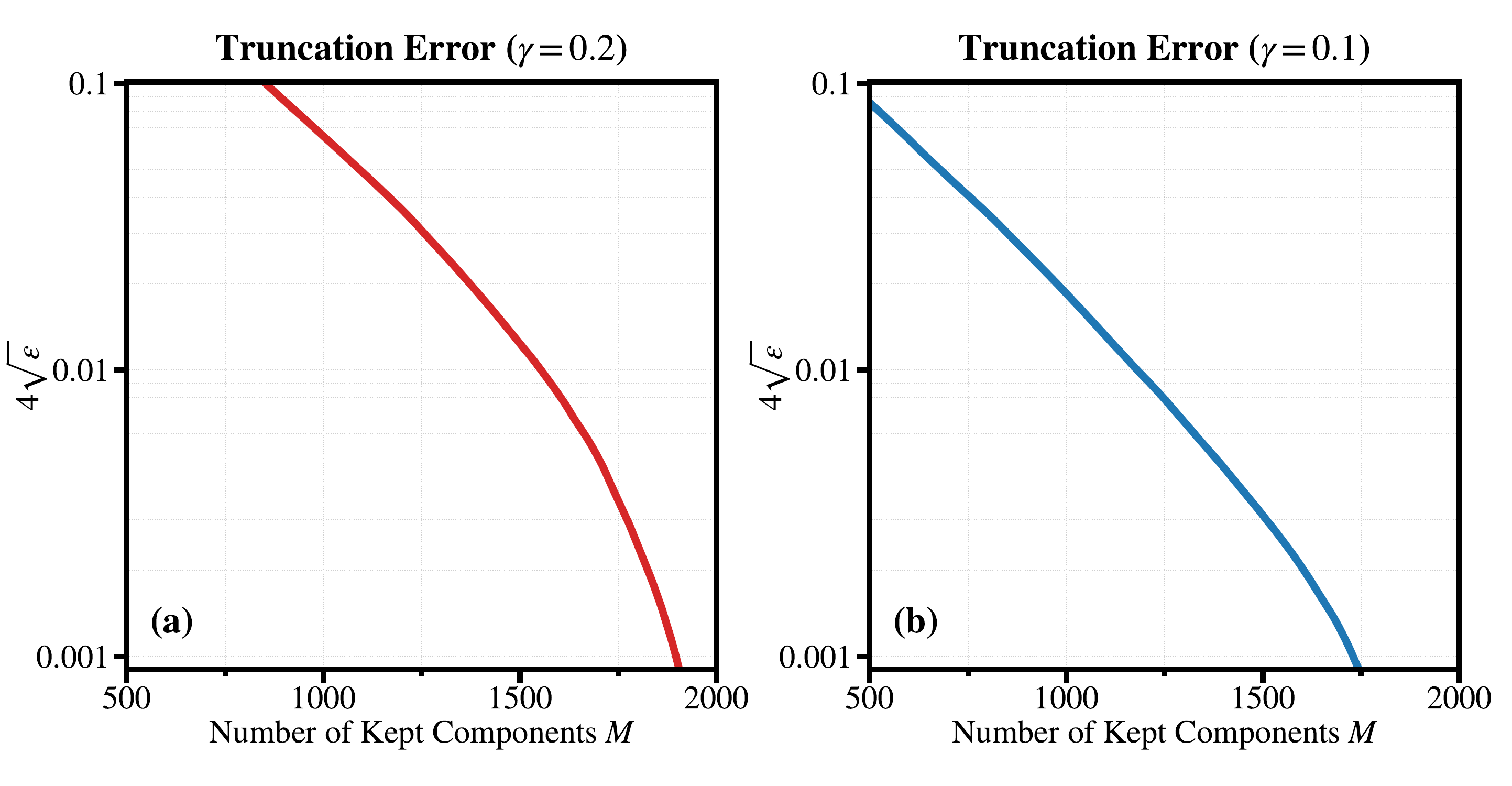}
    \caption{Truncation error $4\sqrt{\epsilon}$ as a function of the number of principal components $M$ for $[[11,1,5]]$ qubit code subject to amplitude damping channels (a) $\gamma = 0.2$. (b) $\gamma=0.1$.}
    \vspace{-5 mm}
    \label{fig:Truncation11}
\end{figure}

Another example is the $[[11,1,5]]$ quantum stabilizer code, whose code space is defined by $10$ cyclic generators:
\begin{align*}
g_1 &= XZZXZIIZIZX,\\
g_2 &= XXZZXZIIZIZ,\\
&\ \vdots \nonumber \\
g_{10} &= ZXZIIZIZXXZ.
\end{align*}
The projector onto the code space is given by
\begin{align}
P_{C} = \prod_{i=1}^{10} \frac{I + g_i}{2}.
\end{align}
Under this construction, the only term in the expansion of $P_{C}$ that yields a non-zero trace in $\Tr(Z_{\boldsymbol{u} \cup \boldsymbol{v}}P_C)$ is the trivial stabilizer $I^{\otimes 11}$. It satisfies the sparse stabilizer configuration assumption. The spectrum concentration is shown in Fig.~\ref{fig:Truncation11}.
\vfill

\end{document}